\documentclass[11pt,a4paper]{article}
\pdfoutput=1
\usepackage{jstyle}

\newcommand{\bi}{\begin{itemize}}
\newcommand{\ei}{\end{itemize}}
\newcommand{\bea}{\begin{eqnarray}}
\newcommand{\eea}{\end{eqnarray}}

\author{Seungho GWAK}
\author{\quad Euihun JOUNG}
\author{\quad Karapet MKRTCHYAN}
\author{\quad Soo-Jong REY}

\affiliation{School of Physics \& Astronomy and Center for Theoretical Physics \\ Seoul National University, Seoul 08826 \rm KOREA}
\affiliation{Gauge, Gravity \& Strings, \ Center for Theoretical Physics of the Universe\\
Institute for Basic Sciences, Daejeon 34047 \rm KOREA}

\emailAdd{epochmaker, 
\ euihun.joung, \ karapet, \ sjrey @ snu.ac.kr}

\title{\centering
\LARGE{Rainbow Vacua \\
of \\
Colored Higher-Spin (A)dS$_3$ Gravity}}

\abstract{We study the color-decoration of higher-spin (anti)-de Sitter gravity in three dimensions. 
We show that the rainbow vacua, which we found recently  for the colored gravity theory, also pertain in the colored higher-spin theory. The color singlet spin-two plays the role of first fundamental form (metric). The difference is that when spontaneous breaking of color symmetry takes place, the Goldstone modes of massless spin-two combine with all other spins and become the maximal-depth partially massless fields of the highest spin in the theory, forming a Regge trajectory.  }

\begin{document}


\maketitle

\rightline{\sl Whenever a theory appears to you as the only possible one,}
\rightline{\sl take this as a sign that you have neither understood the theory}
\rightline{\sl nor the problem which the theory was intended to solve.}
\rightline{Carl Popper, `Objective Knowledge: Evolutionary Approach' (1972)}

\section{Introduction}

In recent years, higher-spin gravity has been studied intensively. In particular,
much progress has been made related to the higher-spin AdS/CFT duality \cite{Klebanov:2002ja,Sezgin:2002rt}. 
In three dimensional theories,  the breakthroughs
took place in several places consecutively: first, the asymptotic symmetry 
of higher-spin gravity has been identified as nonlinear $W$-algebras \cite{Henneaux:2010xg,Campoleoni:2010zq},
then it led to the conjecture of the $W_{N}$ minimal models
as dual CFTs \cite{Gaberdiel:2010pz}. 
Blackhole-like exact solutions were constructed \cite{Ammon:2012wc}.
Most of these results are about the $hs(\l)\oplus hs(\l)$ 
Chern-Simons higher-spin gravity \cite{Blencowe:1988gj,Vasiliev:1992ix}
or the Prokushkin-Vasiliev theory \cite{Prokushkin:1998bq} which contains the former 
as the gauge sector.
Many variant models of three-dimensional higher-spin gravity 
have been considered later on 
and many other interesting features were discovered (see \cite{Afshar:2013vka,Gonzalez:2013oaa,Campoleoni:2012hp,Boulanger:2014vya,
Campoleoni:2014tfa,deBoer:2013vca,Buchbinder:2012xa} for a non-complete list of references).

In the companion paper \cite{CS}, we proposed an extension of the three-dimensional gravity to multi-graviton system --- color-decorated three-dimensional gravity. 
The purpose of this paper is to extend this analysis to  higher spins. 
More precisely, we consider the Chern-Simons formulation of higher-spin (A)dS gravity, 
leaving aside the matter coupling issue of the Prokushkin-Vasiliev theory.
In fact, the possibility of color decoration appears as a
rather natural generalization in the Vasiliev's approach to higher-spin gravity, 
where the higher-spin algebra plays the key role for consistency of the theory and  the color decoration is
one of the simplest extensions of higher-spin algebra. This was first pointed out in \cite{Vasiliev:1986qx,Konstein:1989ij}, and further discussed in \cite{Vasiliev:2004cm}.
Actually, all the Vasiliev's nonlinear higher-spin equations can
be consistently color decorated with the same mechanism. 
Judging from this and also from the dual CFT considerations, 
we anticipate that the issue of color decoration might be more consequential in
the context of higher-spin gravity.

We analyze the colored higher-spin gravity in three dimensions
and show, in particular, that the salient aspect of the colored gravity
 persists in the higher-spin extension:
when the higher-spin gravity is color decorated, there appears a staircase
potential with a number of extrema. Each of these extrema provides 
an (A)dS background solution with a different cosmological constant.
For $SU(N)$ color symmetry, there exist $[\frac{N+1}2]$ different vacua
--- henceforth, referred to as \emph{rainbow vacua} ---
which spontaneously break the color symmetry down to 
$SU(N-k)\times SU(k)\times U(1)$\, ($k=0,1,\dots,[\frac{N-1}2]$).
When this symmetry breaking happens,
the Goldstone modes ---
the spin two fields corresponding to the broken part of the symmetries
--- combine with all other spins
and become a longer spectrum. For the models of higher-spin gravity
involving massless spins up to $M$\,, that is the $\mathfrak{gl}_{M}\oplus
\mathfrak{gl}_{M}$ Chern-Simons model, the symmetry broken part of the spectrum forms
maximal-depth partially massless spin $M$ representation \cite{Deser:2001xr}. 
This representation has as many degrees of freedom (DoF)  as 
the collection of massless spins from 1 to $M$\,.
Therefore, for a generic model of colored higher-spin gravity
--- involving arbitrary higher spins --- we might end up with a rather exotic 
spectrum: maximal-depth partially-massless fields of infinite spin, which is reminiscent of a Regge trajectory. 
In three dimensions, the (partially-)massless fields have only boundary DoF,
but in the limit of infinite spin, the dimension of their phase-space becomes infinite. They correspond in dual conformal field theory living at the boundary to infinite-tower of (partially-)conserved global currents.
We would like to emphasize that the color-decorated higher-spin gravity,
or any variant/generalization of the latter, is 
a plausible bulk theory when considering a bi-vector type free CFT
and looking for its bulk dual in the context of AdS/CFT. 
In this set-up, the rainbow vacua and the spontaneous symmetry breaking 
are generic and unneglectable phenomena.

The organization of the paper is as follows.
In Section \ref{sec: construction}, we review
 how the consistent color decoration works in the
models of higher-spin gravity.
In Section \ref{sec: vacua}, the rainbow vacua 
with different cosmological constants are identified.
In Section \ref{sec: metric}, we expand the theory around 
one of the rainbow vacua by solving the torsionless condition.
In Section \ref{sec: spectrum}, we analyze the spectrum resulting from the symmetry breaking
and show that it corresponds to the maximal-depth partially-massless field.
In Section \ref{sec: rep}, we provide an account for the partially massless representations in three dimensions.
Finally, Section \ref{sec: discussion} contains further discussion. 


\section{Color Decoration of  Higher-Spin (A)dS$_3$ Gravity}
\label{sec: construction}

\subsection{Color Decoration}
We first recapitulate how one can consistently color-decorate a given higher-spin theory, extending our previous work \cite{CS}.

Suppose we are given an uncolored (higher-spin) gravity theory, defined either by an action or by a set of field equations. Assume that elementary fields take values in an (higher-spin) isometry Lie algebra $\mathfrak g_{i}$\,. 
The idea is that in order to color-decorate this theory by  
attaching Chan-Paton factors to the fields,
we require these fields to take values in the tensor product algebra 
$\mathfrak{g}_i \otimes \mathfrak g_c$\,. 
The $\mathfrak{g}_{c}$ is the color symmetry algebra.
However, though we may start with Lie algebras $\mathfrak g_i$ and $\mathfrak g_c$, the tensor product algebra does not automatically provide a Lie algebra $\mathfrak{g}_i \otimes \mathfrak g_c$\, because the anticommutators are not defined. This point should be clear from
\be
	[\,M_{X}\otimes \bm T_{I},M_{Y}\otimes \bm T_{J}\,]
	=\frac12\,[\,M_{X},M_{Y}\,]\otimes \{\,\bm T_{I} ,\bm T_{J}\,\}
	+\frac12\,\{\,M_{X}, M_{Y}\,\} \otimes [\,\bm T_{I},\bm T_{J}\, ]\,.
\label{commutation of product group}
\ee
We conclude that, if an associative product can be defined in $\mathfrak{g}_i$ and $\mathfrak{g}_c$\,, the color-decoration through the Chan-Paton factors can be achieved.

One can always take $\mathfrak{g}_c=\mathfrak{u}(N)$ as color symmetry, and hence the associativity of $\mathfrak{g}_c$ is satisfied.
However, the relevant isometry algebras $\mathfrak{so}(d,2)$ or $\mathfrak{so}(d+1,1)$ for (A)dS$_{d+1}$ space do not have an associative structure for general $d$.
The way out of this problem is to consider a larger algebra $\mathfrak{g}_{i}$ which contains the isometry algebra.
In this way, the $\mathfrak{g}_{i}$ would contain more generators, and so the corresponding uncolored theory would involve more fields than the pure (A)dS$_d$ Einstein gravity.
For instance, in the previous work \cite{CS}, we considered the three-dimensional (anti)-de Sitter gravity, 
where the original isometry algebra $\mathfrak{sl}_{2}\oplus\mathfrak{sl}_{2}$
was first extended to $\mathfrak{g}_{i}=\mathfrak{gl}_{2}\oplus\mathfrak{gl}_{2}$\,.
Apart from (A)dS$_3$ isometries, this algebra contains generators  corresponding
 to two spin-one fields, whose dynamics is described by Chern-Simons action.

The higher-spin theories are particularly suited for the color-decoration,
as discussed earlier in \cite{Vasiliev:1986qx,Konstein:1989ij,Vasiliev:2004cm}. The higher-spin algebra in which the higher-spin fields take values is typically an associative algebra unless one deliberately truncates the theory to the so-called minimal spectrum, containing only spins of even integers. In fact, the color-decoration necessarily requires fields of odd integer spins in the spectrum (spin-one for the pure Chern-Simons (A)dS$_3$ gravity as studied in the previous work \cite{CS}). As such, it is not possible to truncate the spectrum of the colored higher-spin theory to even spins only.

It was also noticed in \cite{Konstein:1989ij} that including fermion generators necessarily requires non-trivial color algebra, therefore realistic models of higher-spin theory in four dimensions should be given by color-decorated theories, possibly with additional color symmetry breaking pattern that leaves only one massless graviton in the spectrum. This is not, however, our concern in this work.

\subsection{Color-Decorated (A)dS$_3$ Higher-Spin Theory}
In this work, we shall consider the simplest class of higher-spin
theory and study their color-decoration. The theory we shall study is the colored version of the Chern-Simons formulation of the higher-spin (A)dS$_3$ theory whose gauge algebra is given by the infinite-dimensional algebra labelled by a continuous parameter $\lambda$:
\be
	\mathfrak{g}_{i}=hs(\lambda)\oplus hs(\lambda)\,.
\ee
To render the conceptual problem simpler, we shall often restrict ourselves to the truncated algebras,
\be
	\mathfrak{g}_{i}=\mathfrak{gl}_{M}\oplus \mathfrak{gl}_{M}\, 
	\qquad (M = 2, 3, 4, \ldots) 
\ee
or even to the simplest higher-spin algebra, the $M=3$ case. The gauge algebra of spin-two, leading to $(A)dS_3$ Einstein gravity, corresponds to $M=2$.  


Let us further discuss aspects of the colored higher-spin (A)dS$_3$ gravity in the Chern-Simons formulation.
The theory is based on the gauge field taking value in $\mathfrak{g}_{i}\otimes\mathfrak{g}_{c}$\,:
\be
	\cA=A^{X,I}\,
	M_{X}
	\otimes \bm T_{I}\,,
	\label{A}
\ee
where $M_{X}$ are the generators of higher-spin algebra  $\mathfrak{g}_{i}$
and the index $\st X$ is the shorthand notation
for the set of indices $\st A_{1}B_{1},\ldots, A_{r}B_{r}$
of higher-spin generators.
The color algebra $\mathfrak{g}_{c}$ is spanned by the generators $\bm{T}_{I}$\,.
The gauge field strength is given by
\be
	\cF= d\cA + \cA\wedge \cA=F^{X,I}\,M_{X}\otimes \bm T_{I}\,.
	\label{F}
\ee
Up to this point, it is clear that all elements of the theory can be straightforwardly color-decorated by adjoining the Chan-Paton indices. Hence,
if a theory can be defined solely in terms of $\cA$ and $\cF$ as in the Chern-Simons formulation --- or
together with more elements which can be equally well color-decorated ---
then the theory can be consistently generalized to a color-decorated version.

The action of the (A)dS$_3$ higher-spin theory is given  in the Chern-Simons formulation by
\be
S =\frac{\k}{4\pi}\, \int 
\tr_{\mathfrak{g}}
\Big(
\cA\wedge d\cA
+\frac{2}{3}\,\cA\wedge \cA\wedge \cA
\Big)\,,
\label{CS}
\ee
where the constant $\k$ is the Chern-Simons level. More concretely, we take the full algebra on which the theory will be based on as
\be
	\mathfrak{g}=\left(
	hs(\lambda)\oplus hs(\lambda)
	\right)
	\otimes \mathfrak{u}(N)
	\ \ominus\ 
	{\rm id}
	\otimes \bm I\,.
	\label{g}
\ee
Note that we subtracted the 
${\rm id}\otimes \bm I$
--- where $\rm id$ and $\bm I$ are the centers
of $hs(\lambda)\oplus hs(\lambda)$ and $\mathfrak{u}(N)$\,, respectively. This generator corresponds to an Abelian Chern-Simons field which does not interact with other fields in the theory. Since gauge field $\cA$ takes value in the subspace of the tensor product space,
the trace $\tr$ of \eqref{CS} should be defined in the 
tensor product space and it is given by the product of two traces as 
$\tr(\mathfrak{g}_{i}\otimes \mathfrak{g}_{c})=
	\tr(\mathfrak{g}_{i})\,\tr(\mathfrak{g}_{c})$\,.

\subsection*{Higher-Spin Algebra}
In the uncolored case, the Chern-Simons theory with
the algebra $hs(\l)\oplus hs(\l)$
can be interpreted as a theory of massless fields
with spins $s=(1), 2,3,4,\ldots$\,,
where spin 1 may or may not be present, depending on 
whether $hs(\l)$ includes the identity or not. 
When the parameter $\l$ takes an integer value, say $M$,
then the $hs(M)$ develops an ideal. The quotient of $hs(M)$ by the
ideal is the finite-dimensional algebra $\mathfrak{gl}_{M}$\,,
whose generators can be organized as
\be
	\mathfrak{gl}_{M}={\rm Span}\{\,J\,,\, J_{a}\,,\, J_{a_{1}a_{2}}\,,\,
	\ldots\,,\, J_{a_{1}\cdots a_{M-1}}\,\}\,,
	\label{gl M}
\ee
whereas the other $\mathfrak{gl}_{M}$ is spanned by $\tilde J_{a_{1}\cdots a_{n}}$\, ($n=0,\dots,M-1$).  $J$ and $\tilde J$ are the identities  
of two copies of $\mathfrak{gl}_{M}$\,.
The basis is chosen in a way 
that makes  manifest that 
the Chern-Simons action with $\mathfrak{gl}_{M}\oplus \mathfrak{gl}_{M}$ algebra
describes a system of massless spins $1,2,3,\ldots, M$\,.

In order to simplify the use of multi indices, let us employ the following notation,
\ba
	&&A_{a(n)}\equiv A_{a\cdots a}
	\quad\leftrightarrow\quad
	A_{a_{1}\cdots a_{n}}\,,\nn
	&&
	A_{a(n)}\,B_{a(m)} 
	\quad\leftrightarrow\quad 
	A_{\{a_{1}\cdots a_{n}}\,B_{a_{n+1}\cdots a_{n+m}\}}\,,
	\label{cv1}
\ea
where the index operation $\{-\}$
means the \emph{traceless} symmetrization:
\ba
	&& A_{\{a_{1}\cdots a_{n}}\,B_{a_{n+1}\cdots a_{n+m}\}}=
	A_{(a_{1}\cdots a_{n}}\,B_{a_{n+1}\cdots a_{n+m})}-({\rm trace})\,,
	\nn
	&&
	\eta^{a_{1}a_{2}}\,A_{\{a_{1}\cdots a_{n}}\,B_{a_{n+1}\cdots a_{n+m}\}}=0\,.
	\label{cv2}
\ea

The algebraic structure of  $hs(\lambda)$ is given by the product,
\be
	J_{a(m)}\,J_{b(n)}
	=\eta_{a(m),b(n)}\,J+\frac1{l!}\,c_{a(m),b(n),c(l)}\,J^{c(l)}\,,
\ee
where $c_{a(m),b(n),c(l)}$ are the structure constants and 
$\eta_{a(m),b(n)}$ is defined by
\be
	\eta_{a(m)}{}^{,b(n)}=\d_{mn}
	\,\big(\d_{a}^{b}\big)^{n}.
\ee
For our analysis, it is not necessary to identify all explicit forms
of $c_{a(m),b(n),c(l)}$\,. It is sufficient to know the following product 
\be
	J_{a(n)}\,J_{b}=
	c_{n}\,\eta_{ba}\,J_{a(n-1)}
	+\epsilon_{ab}{}^{c}\,J_{ca(n-1)}
	+c_{n+1}\,J_{ba(n)}\,,
\ee
where $c_{n}$ can be found e.g in \cite{HS algebra} 
(see \cite{HS algebra recent 1} and references therein for recent works):
\be
	c_{n}=\sqrt{\frac{n(\lambda^2-n^2)}{2n+1}}\,.
	\label{C and D}
\ee
Under the Hermitian conjugation, we get
\be
	(J_{a(n)}, \tilde J_{a(n)})^{\dagger}=
	(-1)^{n}\left\{
	\begin{array}{c}
	(J_{a(n)},\tilde J_{a(n)})\qquad [\s=+1]
	\vspace{3pt}\\
	(\tilde J_{a(n)},J_{a(n)})\qquad [\s=-1]
	\end{array}\right..
\ee

We also define the trace of the identity element as
\be
	\tr(J)=2\,\sqrt{\s}=-\tr(\tilde J)\,. \label{defTr}
\ee
Note that the overall factor 2 of the above trace is 
a matter of convention,
which is related to  the quantization condition of the Chern Simons level $\k$\,.
 It is chosen such that the consistent $\k$ are integers.
We shall come back to this point in Section \ref{sec: Action}.

\section{Color Symmetry Breaking and Rainbow Vacua in General Dimensions}
\label{sec: vacua}

In the previous work \cite{CS},  
we showed that the dynamics of color-decorated (A)dS$_3$ gravity 
is rich enough to trigger a spontaneous color symmetry breaking as the colored spin-two matter fields take nonzero vacuum expectation values proportional to the color-singlet component $g_{\mu \nu}$. Note that, in \cite{CS}, we identified this color-single component as the first fundamental form, viz. the metric. By analyzing the potential of these colored spin-two fields, we identified multiple vacua -- named as \emph{rainbow vacua} --- having different values of cosmological constants.

The existence of panoramic rainbow vacua is 
not a feature unique to the (A)dS$_{3}$ gravity. This feature actually holds true for a generic class of color-decorated (higher-spin) gravity theories 
in any dimensions.
In the following, we shall make this point clear, while keeping  generality of our discussion.

We look for the (A)dS$_{D}$ solutions to the equations of the colored higher-spin gravity, such as Chern-Simons and Vasiliev theory.
As an ansatz, we consider the configuration for which only spin-two components of ${\cal A}$
are non-trivial.
All other components --- the one-forms of the other spins 
and also the zero forms of the Vasiliev's equation --- are set to zero.
This ansatz is actually the only consistent one with the isometry of $(A)dS_D$\,. 
First of all, any odd spin fields cannot take a non-trival vacuum expectation value because there is no odd rank tensor compatible with the $(A)dS_D$ isometry. 
About  even spin fields, one may consider
\be
	\la \varphi_{\mu_1\cdots \mu_{2n}}\rangle
	=v\,g_{\m_1\m_2}\,\cdots\,g_{\m_{2n-1}\m_{2n}}\,,
\ee
where $\varphi_{\mu_1\cdots \mu_{2n}}$ are higher-spin fields
in the Fronsdal description.
However, since  fields of spins greater than four are subject 
to double-traceless constraint,  only scalar and spin-two fields
can take a vacuum expectation value compatible with the $(A)dS_D$ isometry.
Therefore, we take  an ansatz for the one-form gauge field as
\be
	\cA= (\o^{ab}\,\bm I+\bm \Omega^{ab})\,M_{ab}
	+\frac1{\ell}\,(e^{a}\,\bm I+\bm E^{a})\,P_{a}\,,			
\ee
where $\bm \Omega^{ab}$ and $\bm E^{a}$ take values in the Chan-Paton algebra, $\mathfrak{su}(N)$\,.

The idea for finding classical solutions is to require that the configuration does not lead to back-reaction onto the other components of the one-form field, viz. either spin-one or higher-spin (and the zero-form field). It is straightforward to see that these requirements are met if we impose the conditions
\be
	\boldsymbol \Omega^{ab} \wedge
	\boldsymbol \Omega^{cd}\,\big\{M_{ab},M_{cd}\big\}=0\,,
	\quad
	\big\{\,\boldsymbol \Omega^{ab}
	\ \overset{\wedge},\ \bm E^{c}\,\big\}
	\big\{M_{ab},P_c\big\}=0\,,
	\quad
	\bm E^{a} \wedge \bm E^{b}\,
	\big\{P_a,P_b\big\}=0\,.
	\label{commuting A}
\ee
Were if these conditions not met, anti-commutators $\{M_{AB},M_{CD}\}$ would contribute\footnote{We denote by $M_{AB}$ generators of $(A)dS_D$ algebra while $M_{ab}$ and $P_a$ are Lorenz and translation generators, respectively.} and give rise to the generators of fields with 
spin different from two.

We take the ansatz, corresponding to the tensor product structure,  for \eqref{commuting A} as
\be
	\boldsymbol \Omega^{ab}=0\,
	\qquad \mbox{and} \qquad
	\bm E^{a}=e^{a}\,\bm X\,,
\ee
Here, $\bm X$ is a particular element of the $\mathfrak{su}(N)$
to be determined.
Note that we could add a factor in the latter equation but it can be simply absorbed into the matrix $\bm X$\,.
So, our final ansatz for the gauge field takes the form
\be
	\cA = \o^{ab}\,M_{ab}\,\bm I
	+\frac1{\ell}\,e^{a}\,P_{a}\,(\bm I+\bm X)\,.
	\label{ansatz}
\ee
With the above ansatz,
the only non-trivial equation is the  zero curvature equation, 
$\cF=0$\,, which reads
\be
\left(d\omega^{ab}
+\omega^{a}{}_{c}\wedge\omega^{cb}\right)\bm I
+\frac{\s}{\ell^{2}}\,e^{a}\wedge e^{b}\,(\bm I+\bm X)^{2}=0\,,
\qquad
\left( de^a+
\omega^{a}{}_{c}\wedge e^c\right)(\bm I+\bm X)=0\,,		
\label{EOM}
\ee
where $\s$ is a $\pm$ sign, positive for AdS$_{D}$ and negative for dS$_D$.

The first equation in (\ref{EOM}) clearly shows that the Chan-Paton gauge symmetry acts as a new source to the spacetime curvature. To see this more explicitly, let us decompose the above \mt{N\times N} matrix equations into the singlet $\mathfrak u(1)$ part and the $\mathfrak{su}(N)$ part, which in string theory would originate from the {\sl closed string} part and the {\sl open string} part, respectively. The {\sl closed string} part of the equation of motion reads
\ba
	d \omega^{ab}
+\omega^{a}{}_{c}\wedge\omega^{cb}
-\frac{2}{(D-1)(D-2)}\,\L\,e^{a}\wedge e^{b}=0\,,\qquad
	d e^a+\omega^{a}{}_{c}\wedge e^c=0\,,
	\label{cartan}
\ea
where the cosmological constant $\L$
(measured in units of $D$-dimensional Newton's constant) is given by
\be
	\L=-\frac{(D-1)(D-2)\,\s}{2\,N\,\ell^{2}}\,
	\tr(\bm I+\bm X)^{2}\,,
	\label{L d}							
\ee
On the other hand, the {\sl open string} part of the equation of motion is given by
\be
	e^{a}\wedge e^{b} \left(
	2\,\bm X+\bm X^{2}
	-\frac{\tr(2\,\bm X+\bm X^{2})}N\,\bm I\right)=0\,.
	\label{Y eq}								
\ee
We note that, for non-degenerate $e^a$'s, (\ref{Y eq}) is identical to the condition for a critical point of the scalar potential, projected to $\mathfrak{su}(N)$ part:
\be
	V(\bm X)=-\frac{(D-1)(D-2)\,\s}{N\,\ell^{2}}\,
	\tr\left(\bm I+3\,\bm X^{2}+\bm X^{3}\right).
	\label{Pot}
\ee
The cosmological constant in (\ref{L d}) is given by $\L=V(\bm X)/2$ evaluated at extremum points.
Putting \mt{D=3}\,, these are precisely what we found in the
analysis of the color-decorated (A)dS$_3$ gravity in \cite{CS}.
This suggests that the stairwell potential of the color-decorated (A)dS$_3$ gravity
would persist to exist in the generic colored (higher-spin) gravity 
theory in any dimensions.

The complete set of solutions to \eqref{Y eq}
can be found  precisely the same way as in the color-decorated (A)dS$_3$ gravity analyzed in \cite{CS}. 
We simply state the result:
the solution $\bm X$ for \eqref{Y eq} is given by
\be
	\bm X_{k}=\frac{N}{\tr({\bm Z}_{k})}\,\bm Z_{k}-\bm I\,,
\ee
with
\be
	\bm Z_{k}=\begin{bmatrix}
  \bm I_{\sst (N-k)\times(N-k)}   & 0  \\
    0 & -\bm I_{\sst k\times k} 
\end{bmatrix}\,
\label{Z k}
\ee
modulo a $SU(N)$ rotation. Extrema of the potential are labelled by $k=0,1,\ldots, [\frac{N-1}2]$. Moreover, ${\bm X}_k$ at the $k$-th vacuum --- which shifts the gauge field ${\cal A}$ from the $k=0$ (A)dS vacuum to (\ref{ansatz}) --- spontaneously breaks the $\mathfrak{su}(N)$ color symmetry down to $\mathfrak{su}(N-k)\oplus \mathfrak{su}(k)
\oplus \mathfrak{u}(1)$\,. We thus conclude from (\ref{ansatz}) that the vacuum expectation values of colored spin-two fields act as the order parameter of the color symmetry breaking.

\section{Metric Formulation}
\label{sec: metric}

Coming back to our model of colored higher-spin theory in three dimensions 
(\ref{CS}\,,\,\ref{g}),
let us notice that
the appearance of the non-trivial potential with multiple 
extrema and the field contents at such vacua are better treatable in the metric form. The exact expression of the staircase potential can be computed along the same lines as in \cite{CS}, so we shall not aim to repeat the derivation. Rather, we shall focus on the identification of perturbative spectrum around each extremum.

\subsection{Decomposition of Associative Algebra}
Once the frame-like formulation of higher-spin gravity is given, rewriting it 
in the metric form is in principle possible. However, it is technically cumbersome  to get exact expressions in the metric-like variables. 
Here, we shall reformulate the Chern-Simons action to metric form solving the torsionless 
condition for the genuine gravition,
while leaving the other field contents in the first-order formulation.
For this task, it is convenient to decompose the algebra $\mathfrak{g}$ 
\eqref{g} into two pieces $\mathfrak{b}$ and $\mathfrak{c}$\,: 
\be
	\mathfrak{g}=\mathfrak{b}\oplus\mathfrak{c}\,,
	\qquad
	\tr(\mathfrak{b}\,\mathfrak{c})=0\,,
	\label{g=b+c}
\ee
in  a proper way.
The rule of the decomposition is 
that $\mathfrak{b}$ forms a subalgebra 
under which $\mathfrak{c}$ carries an adjoint representation, that is,
\be
	[\mathfrak{b},\mathfrak{b}]\subset\mathfrak{b}\,,
	\qquad
	[\mathfrak{b},\mathfrak{c}]\subset \mathfrak{c}\,.
\ee
Correspondingly to this decomposition of the algebra, 
we also split the one-form gauge field into two parts,
\be
	\cA=\cB+\cC\,,
\ee
where $\cB$ and $\cC$ take values in $\mathfrak{b}$ and $\mathfrak{c}$\, respectively.
In terms of $\cB$ and $\cC$\,,
the Chern-Simons action \eqref{CS} reduces to
\be
	S=
	\frac{\k}{4\pi}\, \int \tr
	\left(
	\cB\wedge d\,\cB
	+\frac{2}{3}\,\cB\wedge \cB\wedge \cB
	+\cC\wedge D_{\cB}\,\cC
	+\frac{2}{3}\,\cC\wedge \cC\wedge \cC
	\right),
\ee
with $D_{\cB}\,\cC=d\,\cC+\cB\wedge\cC+\cC\wedge\cB\,.$
Properly selecting $\mathfrak{b}$ and $\mathfrak{c}$
from the full algebra $\mathfrak{g}$ \eqref{g}\,,
we can conveniently 
handle the manifest covariance with respect to
diffeomorphism and  non-Abelian gauge transformation.

The choice of the decomposition  \eqref{g=b+c}
reflects the symmetry of the background 
around which we are expanding the theory.
Instead of analysing the spectrum separately for the singlet vacuum
and the colored vacua, we directly consider the latter case
since it also covers the former for a special value $k=0$.
In order to begin with the proper decomposition \eqref{g=b+c}, we 
first split the $(A)dS_D$ isometry algebra
(the $\mathfrak{sl}_2\oplus\mathfrak{sl}_2$ subalgebra
of $hs(\l)\oplus hs(\l)$)
 into the Lorentz part $\cM$ and the translation 
part $\cP$ as
\be
	\mathfrak{iso}\simeq \cM\oplus\cP\,.
	\label{iso}
\ee

For the color algebra, 
we take advantage of the fact that, at $k$-th extremum, the  $\mathfrak{su}(N)$ symmetry is broken down to \mt{\mathfrak{su}(k)\oplus \mathfrak{su}(N-k)\oplus \mathfrak{u}(1)}\,.
In accordance with this symmetry breaking, we decompose
the space of the $ \mathfrak{su}(N)$ as
\be
	\mathfrak{su}(N)\simeq
	\mathfrak{su}(N-k)\oplus
	\mathfrak{su}(k)\oplus
	\mathfrak{u}(1)\oplus \mathfrak{bs}\,,
	\label{bs}
\ee
where $\mathfrak{bs}$ is the $2k(N-k)$ dimensional vector space 
corresponding to the \emph{broken gauge symmetry} generators. 
Then, the background matrix $\bm Z_{k}$ \eqref{Z k}
enjoys either commutation or anti-commutation properties
with each of these generators:
\be
	\big[\,\bm Z_{k}\,,\,\mathfrak{su}(N-k)\oplus
	\mathfrak{su}(k)\oplus \mathfrak{u}(1)\,\big]=0\,,
	\qquad
	\big\{\,\bm Z_{k}\,,\,\mathfrak{bs}\,\big\}=0\,.
\ee
Taking advantage of these two decompositions \eqref{iso} and \eqref{bs}
of $\mathfrak{iso}$ and $\mathfrak{g}_{c}$\,,
we now decompose the full algebra $\mathfrak{g}$ \eqref{g} according to 
\eqref{g=b+c}\,:
the gravity plus gauge sector $\mathfrak{b}$ and 
the matter sector $\mathfrak{c}$\,.

The gravity plus gauge sector  has two parts $\mathfrak{b}=\mathfrak{b}_{\rm GR}
	\oplus \mathfrak{b}_{\rm Gauge}$\,:
\be
	\mathfrak{b}_{\rm GR}=
	\big(\cM\otimes{\bm I}\big)
	\oplus
	\big(\cP\otimes \bm Z_{k}\big)\,,
	\qquad
	\mathfrak{b}_{\rm Gauge}=
	{\rm id}\otimes 
	\Big(\mathfrak{su}(N-k)\oplus \mathfrak{su}(k)\oplus \mathfrak{u}(1)\Big)\,.
	\label{b gr gauge}
\ee
In the gravity sector, the isometry algebra is deformed by $\bm Z_{k}$ as
\be
	\bm M_{ab}=M_{ab}\,\bm I\,,
	\qquad
	\bm P_{a}=P_{a}\,\bm Z_{k}\,,
	\label{mat M P}
\ee
but  still satisfies the same commutation relations 
as the undeformed one.
We now specify the fields corresponding to the $\mathfrak{b}$ sector as 
\be
	\cB_{\rm GR}=
	\frac12\left(\o^{ab}+\O^{ab}\right)\bm M_{ab}
	+\frac1{\ell_{k}}\,e^{a}\,\bm P_{a}\,,
	\qquad
	\cB_{\rm Gauge}=
	\bm A_{+}+\bm A_{-}
	+\tilde{\bm A}_{+}+\tilde{\bm A}_{-}+(A+\tilde A)\,\bm Y_k\,,
	\label{B k}
\ee
where the traceless matrix
\be
 \bm Y_{k}=\frac{k\,\bm I_{+}
	-(N-k)\,\bm I_{-}}N\,,
\ee
corresponds to $\mathfrak{u}(1)$ symmetry.
The tensor $\O^{ab}$ has been introduced so that 
the spin connection $\o^{ab}$ is 
determined only by $e^{a}$ when solving the torsionless condition. Consequently, the $\O^{ab}$ contains
 the contributions from other matter and higher-spin fields.
The gauge fields $\bm A_{\pm}$ and $\tilde{\bm A}_{\pm}$ take value 
in $\mathfrak{su}(N-k)$ for the subscript $+$ and 
$\mathfrak{su}(k)$ for the subscript $-$\,. 
Finally, the deformed radius $\ell_{k}$ is related to the undeformed one as
\be
	\ell_{k}=\frac{N-2k}{N}\,\ell\,.
\ee

The matter sector $\mathfrak{c}$ has four parts:
\be
	\mathfrak{c}=\mathfrak{c}_{\rm CM}
	\oplus \mathfrak{c}_{\rm BS}
	\oplus \mathfrak{c}_{\rm HS}\,.
\ee
For the introduction of each element, we need to define first the deformed higher-spin algebra generators
analogously to the spin-two sector \eqref{mat M P} as 
\be
	\bm J_{a(n)}=J_{a(n)}\,
	\bm I_{+}+\tilde J_{a(n)}\,\bm I_{-}\,,
	\qquad
	\tilde{\bm J}_{a(n)}
	=J_{a(n)}\,\bm I_{-}
	+\tilde J_{a(n)}\,\bm I_{+}\,,	
	\label{twisted HS J}
\ee
where $\bm I_{\pm}$ are the identities associated with
$\mathfrak{u}(N-k)$ and $\mathfrak{u}(k)$\,, respectively:
\be
	\bm I_{\pm}=\frac12\left(\bm I\pm\bm Z_{k}\right).
\ee
Analogous to the deformation of the spin-two part \eqref{mat M P}\,,
the deformed higher-spin generators \eqref{twisted HS J}
still form $hs(\l)\oplus hs(\l)$ algebra.
In terms of these generators, we define the one form fields 
corresponding to the colored matter  $\cC_{\rm CM}$ and the color-neutral matter $\cC_{\rm NM}$ as
\ba
\cC_{\rm CM}\eq\frac1{\ell_{k}}
	\sum_{n\ge1}\frac1{n!}\Big[\left(\bm\varphi^{a(n)}_{+}
	+\bm\varphi^{a(n)}_{-}\right)
	\bm J_{a(n)}
	+\left(\tilde{\bm\varphi}^{a(n)}_{+}
	+\tilde{\bm\varphi}^{a(n)}_{-}\right)
	\tilde{\bm J}_{a(n)} \Big] \\
\cC_{\rm NM} \eq {1 \over \ell_k} 
\sum_{n\ge1} {1 \over n!} 
	\left(\psi^{a(n)}\,\bm J_{a(n)}
	+\tilde\psi^{a(n)}\,\tilde{\bm J}_{a(n)}\right)
	 \bm Y_{k}\,\bm Z_k \,,	
	\label{C NM}
\ea
where the fields $\bm\varphi^{a(n)}_{+/-}$ and 
$\tilde{\bm\varphi}^{a(n)}_{+/-}$ take value 
in $\mathfrak{su}(N-k)/\mathfrak{su}(k)$
and $\psi^{a(n)}$ and $\tilde\psi^{a(n)}$ in $\mathfrak{u}(1)$\,.
The last matrix factor $\bm Y_{k}$ has been introduced so that
$\tr(\mathfrak{b}_{\rm GR}\,\mathfrak{c}_{\rm CM})=0$. Equivalently,
\be
	\tr\big(\bm J\,\bm Y_{k}\bm Z_k\big)
	=0
	=\tr\big(\tilde{\bm J}\,\bm Y_{k}\bm Z_k\big)\,.
\ee 
The one-form corresponding to the $\mathfrak{c}_{\rm BS}$ sector 
is given by
\ba
	\cC_{\rm BS}\eq\frac1{\ell_{k}}
	\sum_{n\ge0}\frac1{n!}\left(
	\bm\phi^{a(n)}\,\bm J_{a(n)}
	+\tilde{\bm\phi}^{a(n)}\,\tilde{\bm J}_{a(n)}\right),
\ea
where  $\bm\phi^{a(n)}$ and $\tilde{\bm\phi}^{a(n)}$ take values in $\mathfrak{bs}$ \eqref{bs}.
Notice that the summation starts from $n=0$
so it involves not only higher-spin generators but also the identity piece
corresponding to spin one.
Lastly, we have the singlet higher-spin sector:
\be
	\cC_{\rm HS}=\frac1{\ell_{k}}
	\sum_{n\ge2}\frac1{n!}\left(
	\varphi^{a(n)}\,\bm J_{a(n)}
	+\tilde{\varphi}^{a(n)}\,\tilde{\bm J}_{a(n)}\right),
\ee
which, in principle, could be treated together with the
gravity plus gauge sector. 
But, since we do not know any natural form of higher-spin 
covariant interactions in metric-like form, 
they are treated here as extra matter fields.

\subsection{Action in Metric Formulation}
\label{sec: Action}

Putting all the above results into the Chern-Simons action, we get
\be
	S=S_{\rm CS}+S_{\rm HSG}+S_{\rm Matter}\,,
\ee
where the first term $S_{\rm CS}$ is 
the two copies of the Chern-Simons action with the levels $\k$ and $-\k$, respectively, and with $\mathfrak{su}(N-k)\oplus \mathfrak{su}(k)\oplus \mathfrak{u}(1)$ gauge algebra.
In the uncolored Chern-Simons (higher-spin) gravity, 
it is not clear whether the level has to be quantized because the gauge group is  non-compact. In the color-decorated cases,
the level $\k$ ought to take a discrete value 
for the consistency of large color gauge transformations.

The second term $S_{\rm HSG}$ 
is the action in metric form --- only for the gravity part --- for higher-spin gravity,	
\be
	S_{\rm HSG}=
	\frac1{16\pi\,G}\int 
	\left[d^{3}x\sqrt{|g|}\left(R+\frac{2\,\s}{\ell_k^{2}}\right)
	+ L(\varphi,\tilde\varphi,\ell_{k})\right],
	\label{HSG}
\ee
where $L$ is for single higher-spin fields, given by
\ba
	&&L[\varphi,\tilde{\varphi},\ell]
	= L_{+}[\varphi,\ell]-L_{-}[\tilde{\varphi},\ell]\,,\nn
	&&L_{\pm}[\varphi,\ell]=2 \sqrt{\sigma} \sum_{n}\frac 1 {n!}\,
	\bigg[\,
	\frac1\ell\,\varphi^{a(n)}\wedge 
	\left(D\,\varphi_{a(n)}
	\pm\frac{1}{\sqrt{\s}\,\ell}\,
	c_{a(n)bc(n)}\,e^{b}\wedge\varphi^{c(n)}\right)\nn
	&&\hspace{90pt}
	+\,\frac{2}{3\,\ell^{2}}\,
	c_{a(m)b(n)c(l)}\,\varphi^{a(m)}
	\wedge\varphi^{b(n)}
	\wedge\varphi^{c(l)}\,\bigg]\,.
	\label{L}
\ea
The  derivation $D$ is covariant both with respect to Lorentz transformations and non-Abelian gauge transformations (in case it acts on color charged fields).
At quadratic order, components with different $n$
are independent and describe massless spin $(n+1)$\,.
The gravitational constant $G$ is fixed in terms of
the Chern-Simons level by
\be
	\k=\frac{\ell}{4\,N\,G}\,.
\ee

Finally, the matter action takes the form,
\ba
	S_{\rm Matter}\eq
	\frac{1}{16\,\pi\,G}\int 
	\frac{1}{N-2k}\,\tr\Big(
	L[\bm\varphi_{+},\tilde{\bm\varphi}_{+},\ell_{k}]
	-L[\bm\varphi_{-},\tilde{\bm\varphi}_{-},\ell_{k}]\Big)\nn
	&&\hspace{45pt}
	-\,\frac{k(N-k)}{N^{2}}\,L[\psi,\tilde{\psi},\ell_{k}]
	+\frac{1}{N-2k}\,L_{\rm\sst BS}[\bm\phi,\tilde{\bm\phi},\ell_{k}]
	+L_{\rm int}\,,
	\label{Matter k}
\ea
where the new three form $L_{\rm\sst BS}$
has fully correlated components as opposed to $L$ \eqref{L}\,:  
\ba
	&& L_{\rm\sst BS}[\bm\phi,\tilde{\bm\phi},\ell] \label{L pm} \\
	&=&\frac{4\,\sqrt{\s}}{\ell}\,
	\sum_{n}\frac1{n!}\,\tr
	\left[\tilde{\bm\phi}_{a(n)}\wedge 
	\left(D\,\bm\phi^{a(n)}+\frac{c_{n}}{\sqrt{\s}\,\ell}\,e_{a}\wedge\bm\phi^{a(n-1)}
	+\frac{c_{n+1}}{\sqrt{\s}\,\ell}\,	e_{a}\wedge \bm\phi^{a(n+1)}\right)
	 \bm Z_{k}\right]. \nonumber	
\ea
The term $L_{\rm int}$ in the second line of \eqref{Matter k} 
concerns exclusively
the interaction terms and it
contains the cross couplings from the Chern-Simons cubic interaction
and the quadratic terms in $\Omega^{ab}$ (which
itself is quadratic in fields, hence these terms represent quartic couplings).
Note that, in $k$-th vacua, the colored matter interacts with other fields with strength proportional to powers of $\sqrt{N-2 k}$, while the neutral matter $\psi, \tilde{\psi}$ interact with strength proportional to powers of $1/\sqrt{k (N - k)}$. 


In describing the action of colored higher-spin gravity above, 	
we omitted the explicit expression for the structure constants $c_{a(m),b(n),c(l)}$\,, $L_{\rm int}$ and $\O^{ab}$\,.
Identifying their form is straightforward in principle
but not necessary for our purpose:
we are more interested in the pattern of mass spectra around rainbow vacua and the qualitative structure of interactions.
In the following, we elaborate more on these aspects.
\begin{itemize}
\item
The spectrum consists of spin-one Chern-Simons gauge fields,
spin-two gravity, and higher-spin fields,
whose dynamics are governed by 
$S_{\rm CS}$ and $S_{\rm HSG}$\,.
In particular, the latter $S_{\rm HSG}$ coincides with the action of the uncolored Chern-Simons higher-spin theory in three-dimensions.
The colored higher-spin fields $\bm\varphi_{\pm}^{a(n)}$ and singlet higher-spin fields $\psi^{a(n)}$
(and their tilde counter parts)
share the same structure of the quadratic Lagrangian $L$ \eqref{L}. As such, they describe massless spin-$(n+1)$ fields, described in first-order formalism.

\item From the signs of the one-derivative term and masslike term in the first-order Lagrangian \eqref{L}\footnote{In $(2+1)$ dimensions,  the sign of one-derivative term in first-order Lagrangian signifies `chirality' , while the sign of zero-derivative 'mass' term determines (non)unitarity.  Details of this peculiar feature is explained in our companion paper \cite{CS}, Section 5.2 and Appendix A.}, 
it follows that the colored matter $\bm\varphi_{+}^{a(n)}, \tilde{\bm\varphi}_{+}^{a(n)}$ are unitary, but $\bm\varphi_{-}^{a(n)}, \tilde{\bm\varphi}_{-}^{a(n)}$ and singlet matter $\psi^{a(n)}, \tilde\psi^{a(n)}$ in \eqref{Matter k}
behave as non-unitary ghosts. This reflects that all of the $0  < k < N/2$ rainbow vacua are actually
saddle points and so they have unstable, runaway directions in the field configuration space.

\item The remaining bi-fundamental matter fields $\bm\phi^{a(n)}$ and $\tilde{\bm\phi}^{a(n)}$
corresponds to the broken part of the color symmetries. They are all massive. 
More precisely, as we shall demonstrate in detail in the next section, these fields are  the partially-massless fields  \cite{Deser:2001xr} of maximal-depth.

\item
The color symmetry breaking mechanism through colored higher-spin gravity should be contrasted against the more familiar standard Higgs mechanism through colored scalar field dynamics.
The role of the Higgs field and its symmetry-breaking potential is 
now played by the $\mathfrak{su}(N)$-valued spin-two matter field
and its symmetry-breaking potential $V(\bm X)$ \eqref{Pot}. Much as colored spin-0 field condensate respects the Poincar\'e invariant vacuum in the former, colored spin-2 field condensate respects the generally covariant vacuum in the latter.  When this field takes a nonzero vacuum expectation value corresponding to one of the saddle points (labeled by $k$ as in \eqref{Z k}),
its components split into two parts, each of which retains the residual
$\mathfrak{su}(N-k)\oplus\mathfrak{su}(k)\oplus \mathfrak{u}(1)$ symmetry.
The symmetry preserving part remains massless,
while the symmetry breaking part
--- analogous to the Goldstone bosons --- 
combine with the same part
of all other higher-spins.
Hence, the massless spin-2 Goldstone field
combines with masless fields of spin $1,3,4,\ldots, M$ in case 
of  the $\mathfrak{gl}_M\oplus \mathfrak{gl}_M$ higher-spin gravity.
The resulting spectrum is the spin-$M$ partially-massless fields of maximal-depth. 
We expect the same pattern continues to hold for $hs(\lambda) \oplus hs(\lambda)$ higher-spin gravity:
the massless colored spin-2 Goldstone field combines to massless colored fields of all other spins and form a partially massless Regge trajectory.

\item
The interaction among the above fields is set not only by the gravitational constant but also by $k, N$.  
The structure of interaction in the color non-singlet vacua
is analogous to the colored (A)dS$_3$ gravity we studied in the previous work \cite{CS}.
All the fields are coupled to gravity in the diffeomorphism invariant manner.
All the colored ihgher-spin fields ---
adjoints $\bm\varphi_{\pm}$ and bi-fundamentals $\bm\phi$ ---
have covariant gauge couplings to the Chern-Simons gauge fields.
There are also nonlinear self-couplings among the matter fields with coupling constants controlled by $N$ and $k$ \eqref{Matter k}. 
These nonlinear interactions become strong for small $k$ (small symmetry breaking)
and as weak as gravity for large $k \sim [N/2]$ (large symmetry breaking).
\end{itemize}

\section{Mass Spectrum of the Broken Color Symmetries}
\label{sec: spectrum}

We already noted that the color symmetry breaking triggers the mass generation as well. Moreover, it suggested a mechanism for emergent Regge trajectory out of massless higher-spin fields.  This is an important aspect by itself, so we analyze below the spectrum of these ``massive'' components.

\subsection{General Structure} 
We now analyze the action for the  bi-fundamental higher-spin fields $\bm\phi^{a(n)}$ 
and $\tilde{\bm\phi}^{a(n)}$ corresponding
to the broken part of color symmetries, paying special attention to their mass spectra. For definiteness, we concentrate on the AdS space. 
To get analogous result for dS space, we simply replace the AdS radius to 
$\sqrt{-1}$ times the dS radius.

It turns out that all these fields with different $n$ are correlated.
Furthermore, even the left-movers and the right-movers have cross-couplings in the quadratic action. However, we can always diagonalize the acton. Taking, for clarity of the analysis, the $\mathfrak{gl}_M$ higher-spin algebra,  we can reduce the action to a collection of $S_{\rm\sst BS}$ given by
\ba
	&& S_{\rm\sst BS}[\phi,\phi^{a},\ldots,\phi^{a(M-1)}]=\nn
	&&\qquad
	=\,\sum_{n=0}^{M-1} (-1)^n \int 
	\phi_{a(n)}\wedge 
	\left[
	D\,\phi^{a(n)}
	+\frac{ 1}{\ell}\left(
	c_{n}	\,e^{a}\wedge \phi^{a(n-1)}
	+c_{n+1}\,	e_{a}\wedge \phi^{a(n+1)}\right)\right].\ 
		\label{PM HS action}
\ea
Notice that the one-form fields contributing to the action
are truncated to the first  $M$ fields.
The above action also admits gauge symmetries with parameters 
$(\varepsilon,\varepsilon^{a},\ldots,\varepsilon^{a(M-1)})$ as
\be
	\delta\,\phi^{a(n)}
	=D\,\varepsilon^{a(n)}
	+\frac {1 }{\ell}\left(
	c_{n}	\,e^{a}\,\varepsilon^{a(n-1)}
	+c_{n+1}\,	e_{a}\,\varepsilon^{a(n+1)}\right).
	\label{gl M gauge}
\ee

For the analysis of equations of motion, 
we consider the decomposition of $\phi^{a(n)}$ into 
\ba
	\bar h_{\mu(n+1)}
	\eq \big(e_{\mu}{}^{a}\big)^{n}\,
	\phi_{\mu\,a(n)}\,,\nn
	h'_{\mu(n-1)}
	\eq \big(e_{\mu}{}^{a}\big)^{n-1}\,e^{\m b}\,
	\phi_{\mu\,a(n-1)b}\,,\nn
	f_{\mu(n),\nu}
	\eq \big(e_{\mu}{}^{a}\big)^{n}\,\phi_{\nu\,a(n)}
	+
	e_{\nu}{}^{a}\,
	\big(e_{\mu}{}^{a}\big)^{n-1}\,
	\phi_{\mu\,a(n)}\,,
	\label{field decom}
\ea
where $\bar h_{\m(n+1)}$ and $h'_{\m(n-1)}$ are 
totally symmetric traceless
fields  and $f_{\mu(n),\nu}$'s are
the traceless fields of the Young-symmetry type $\{n,1\}$\,.
Note that we are using the same 
repeated index convention as \eqref{cv1} and \eqref{cv2}.

The procedure of the analysis  can be 
summarized in the following steps:
\begin{itemize}
\item
We first gauge-fix 
$\bar h_{\m(n)}$ from $n=1$ to $M-1$
using the gauge transformations \eqref{gl M gauge} 
with the parameters $\varepsilon_{\m(n)}$ from $n=1$ to $M-1$\,.
\item
Using the equations of motions, 
all the hook fields $f_{\mu(n),\n}$ can be algebraically determined 
in terms of the rest.
At this stage, the residual field contents are
\be
	\bar h_{\mu(M)}\,,
	\qquad
	h'_{\mu(n)}\quad [n=0,\ldots,(M-2)]\,,
\ee
and these fields combine to form two traceful fields of
 spin $M$ and $M-3$\,, respectively.
This is the field content of massive higher-spin fields along the lines taken by Singh and Hagen 
\cite{Singh},
except that, in our case, we also have a gauge symmetry
with the scalar parameter $\varepsilon$\,.
This already suggests that the spectrum described by this system
corresponds to the maximal-depth partially-massless spin-$M$ field\,.

\item
Other equations can be used to algebraically
determine $h'_{\mu(n)}$ from $n=1$ to $(M-2)$\,.
Hence, after this step, we end up only with $\bar h_{\mu(M)}$
and $h' \equiv h'_{\mu(0)}$, modulo the gauge equivalence 
given by the scalar parameter $\varepsilon$\,.
In the $M=2$ case, $\bar h_{\m\n}$ and $h'$ can combine
to a single traceful field $h_{\m\n}$\,.

\end{itemize}

The final equation is of first-order type 
and involves the fields $\bar h_{\mu(M)}$ and 
$h'$\,.
These fields have gauge symmetries
involving $M$ derivatives for $\bar h_{\mu(M)}$
and of second-order for $h'$\,. 
To analyze further, instead of proceeding with the generic value of $M$\,,
we shall consider the $M=3$ example in detail.
The analysis for generic values of $M$ is a straightforward generalization
and they will be 
presented in a forthcoming paper \cite{forthcoming} along  with the analysis of the colored Vasiliev's equations.

\subsection{Example: $\mathfrak{gl}_3\oplus\mathfrak{gl}_3$}

For more concrete understanding, let us explicitly analyze 
the $M=3$ case. From \eqref{field decom}, we get seven fields 
\be
	\bar h_{\m\n\r}\,,\quad
	\bar h_{\m\n}\,,\quad
	\bar h_{\mu}\,,\quad
	f_{\m\n,\r}\,,\quad
	f_{\m\n}\,,\quad
	h'_{\m}\,,\quad
	h'\,.
\ee
They admit the equations of motion,
\ba
	&&
	\nabla_{[\m}\,\bar{h}_{\n]}{}^{\r\s}
	+\frac{4}{3}\,\nabla_{[\m}\,f^{\r\s}{}_{,\n]} 
	-\frac{3}{5}\,\delta_{[\m}^{\{\r}\,\nabla_{\n]}\,h'^{\s\}}
	+\frac{2\sqrt{2}}{\ell}\, \delta_{[\m}^{\{\r}\,\bar{h}_{\n]}{}^{\s\}}=0\,,\\
	&&
	\nabla_{[\m}\bar{h}_{\n]\r}
	+\nabla_{[\m}f_{\n]\r}
	+g_{\r[\m}
	\left(
	\frac{1}{\ell}\,\bar{h}_{\n]}
	-\frac{1}{3}\,\partial_{\n]}h'
	\right)=\frac{2\sqrt{2}}{3\,\ell} 
	\left(f_{\r[\n,\m]}-\frac{3}{4}\,g_{\r[\n}\,h'_{\m]}
	\right),
			\label{algebraic 1}
	\\
	&&
	\partial_{[\m}\bar{h}_{\n]}
	=\frac{8}{3\,\ell}\,f_{\m\n}\,,
		\label{algebraic 2}
\ea
where \eqref{algebraic 1} and \eqref{algebraic 2} simply imply 
that $f_{\m\n,r}$, $f_{\m\n}$ and $h'_{\mu}$ 
are determined by the rest.
The remaining fields have the gauge symmetries,
\ba
	\delta \bar h_{\m\n\r}
	=\nabla_{\{\m}\,\varepsilon_{\n\r\}}
	\,,&\qquad
	& \delta \bar h_{\m\n}
	=\nabla_{\{\m}\,\varepsilon_{\n\}}
	+\frac{1}{\sqrt{2}\,\ell}\,\varepsilon_{\m\n}
	\,,
	\nn
	\delta \bar h_{\mu}=\partial_{\m}\,\varepsilon 
	+\dfrac{8}{3\,\ell}\,\varepsilon_{\m}
	\,,&
	\qquad
	&\delta h'=\nabla^\r\varepsilon_\r+\frac{3}{\ell}\, 
	\varepsilon\,.
\ea
One can first gauge fix $\bar h_{\m\n}$ and $\bar h_{\m}$ 
using the gauge transformations 
with the parameters $\varepsilon_{\m\n}$ and $\varepsilon_{\n}$\,.
 This gauge fixing will relate the latter gauge parameters to 
the scalar one $\varepsilon$ as
\be
	\varepsilon_{\m\n}=\frac{3\,\ell^2}{4\sqrt{2}}\,
	\nabla_{\{\m}\partial_{\n\}}\,\varepsilon
	\,,\qquad
	\varepsilon_{\m}=-\frac{3\,\ell}{8}\,\partial_\m\,
	\varepsilon
	\,.
\ee
Finally, the remaining equations of motions and gauge
transformations are given by
\be
	\nabla_{[\m}\,\bar h_{\n]\r\s}
	+\frac{\sqrt{2}\,\ell}{5}\,\nabla_{[\m}\,g_{\n]\{\r}\,\nabla_{\s\}}\,h'=0\,,
\ee
and
\be
\delta \bar h_{\m\n\r}
=\ell^{2}\ \nabla_{\{\m}\!\nabla^{\phantom{a}}_{\n}\nabla_{\r\}}
\,\varepsilon
\,,
\qquad
\delta h'
=  -
\frac{\ell}{\sqrt{2}}
\left(\nabla^{2}- \frac{8}{\ell^2}\right)\varepsilon\,.
\ee
These gauge transformations precisely coincide 
with those of the maximal-depth partially-massless field,
which has been studied e.g. in \cite{Deser:2001xr} and \cite{Zinoviev:2001dt}.
Hence, this $M=3$ example demonstrates that the spectrum 
corresponding to the broken part of color symmetry is indeed
the maximal-depth partially-massless fields of the highest spin in the theory.

\section{Partially-Massless Fields in Three Dimensions}
\label{sec: rep}

A novel aspect of the color-decorated (higher-spin) (A)dS$_3$ gravity is that the fields in the broken symmetry sector, which acquired masses via Higgs mechanism through the color symmetry breaking, are all partially-massless. In this section, we discuss salient features of these states in (A)dS$_3$ space. 
 
Partially-massless fields carry irreducible representations of the isometry algebra of non-vanishing constant curvature background~\cite{Deser:1983tm, Deser:1983mm, Deser:2001us}\footnote{For framelike formulation of partially-massless higher-spin fields, see \cite{Skvortsov:2006at}. For early discussion on (A)dS/CFT correspondence of partially-massless fields, see \cite{Dolan:2001ih}. }. 
In dS space, these states are unitary. In AdS space, even though their energy is bounded from below, these states are nonunitary because of negative norm states involved. For a given spin $s$\,, there are $s$ different partially-massless fields labelled by depth $t=0,1,\ldots, s-1$, where $t=0$ case corresponds to the massless 
field. In the flat limit, depth $t$ partially-massless field is reduced to a set of massless fields with helicities $s, s-1, \ldots, s-t$\,.
This pattern manifests the number of degrees of freedom (DoF) they have 
\cite{Deser:2001xr}: 
the number interpolates between those of massless and massive fields. In the case of minimal-depth with $t = 0$, we already mentioned that the partially-massless field is a massless spin-$s$ field.  
In the case of maximal-depth with $t=s-1$\,, the partially-massless field contains just one less DoF --- corresponding to a scalar field --- compared to a massive spin-$s$ field.

\subsection*{$AdS_3$ case}
Let us first consider AdS case $\mathfrak{so}(2,d)$
and its lowest-weight representation $\cV_{\mathfrak{so}(2,d)}(\D,s)$
 labeled by the lowest energy $\D$ and spin $s$\,. 
The unitarity bound $\D=s+d-2$ corresponds to the massless field (for $s\ge1$) and the depth $t$ partially-massless fields corresponds to $\D=s+d-2-t$\,. In these cases, we have to factor out invariant subspaces corresponding to gauge modes in order to describe irreducible representations .

In three dimensions, 
 the lowest-weight representation $\cV_{\mathfrak{so}(2,2)}(\D,s)$ is decomposed into those of 
two $\mathfrak{so}(2,1)$ in
$\mathfrak{so}(2,2)\simeq\mathfrak{so}(2,1)\oplus\mathfrak{so}(2,1)$ as
\be
	\cV_{\mathfrak{so}(2,2)}(h_{1}+h_{2},h_{1}-h_{2})=
	\big[\,\cV_{\mathfrak{so}(2,1)}(h_{1})\otimes 
	\cV_{\mathfrak{so}(2,1)}(h_{2})\,\big]
	\oplus \big[\,\cV_{\mathfrak{so}(2,1)}(h_{2})\otimes 
	\cV_{\mathfrak{so}(2,1)}(h_{1})\,\big]\,,
\ee
where we identify the lowest-weights and spin of $\mathfrak{so}(2,2)$
with those of $\mathfrak{so}(2,1)$'s
as $\D=h_{1}+h_{2}$ and $s=|h_{1}-h_{2}|$\,.
Here, we focus on the parity-invariant representations, and so include both of the $\pm s$ helicities  assuming $s\neq 0$\,.
If $s=0$\,, then we simply get 
$\cV_{\mathfrak{so}(2,2)}(2h,0)=
	\cV_{\mathfrak{so}(2,1)}(h)\otimes 
	\cV_{\mathfrak{so}(2,1)}(h)$\,.
In terms of $V_{h}:=\cV_{\mathfrak{so}(2,1)}(h)$\,, it is simpler to understand the appearance of invariant subspace:
when $h$ takes a non-positive half-integer value,
the representation splits into
\be
	V_{-h}=R_{h}\oplus 
	V_{h+1}\qquad [\,2h\in \mathbb N\,]\,,
\ee
where $V_{h+1}$ is
the infinite-dimensional invariant subspace 
and $R_{h}$ is the $(2h+1)$-dimensional 
representation.

Now considering the lowest-weight of partially-massless fields, $\D=s-t$\,,
the lowest-weight representation $\cV_{\mathfrak{so}(2,2)}(\D,s)$ reduces to
\ba
	\cV_{\mathfrak{so}(2,2)}(s-t,s)
	\eq
	\left(\,V_{s-\frac t2}\otimes V_{-\frac t2}\,\right)
	\oplus  \left(\,V_{-\frac t2}\otimes V_{s-\frac t2}\,\right)
	\nn
	\eq
	\left[\,V_{s-\frac t2}\otimes
	\left(\,R_{\frac t2}\oplus V_{\frac t2+1}\,\right)\,\right]
	\oplus
	\left[\,\left(\, R_{\frac t2}\oplus V_{\frac t2+1}\,\right)
	\otimes V_{s-\frac t2}\,\right],
	\label{all modes}
\ea
which involve an invariant subspace, 
\be
	\cV_{\mathfrak{so}(2,2)}(s+1,s-t-1)=
	\left(\,V_{s-\frac t2}\otimes V_{\frac t2+1}\,\right)
	\oplus \left(\,V_{\frac t2+1}\otimes V_{s-\frac t2}\,\right)\,,
	\label{gauge mode}
\ee
corresponding to the gauge modes.
After factoring out this, the remaining representation
corresponds to the partially-massless ones:
\be
	\cD_{\mathfrak{so}(2,2)}(s-t,s)=
	\left(\,V_{s-\frac t2}\otimes R_{\frac t2}\,\right)
	\oplus \left(\,R_{\frac t2}\otimes V_{s-\frac t2}\,\right).
	\label{boundary dof}
\ee

Distinct from the massless
case where $R_{0}$ is the trivial representation, due to $R_{t/2}$\,, the partially-massless field cannot be decomposed neatly into left-moving and right-moving (or holomorphic and anti-holomorphic) parts .
Moreover, $R_{t/2}$ is the finite-dimensional representation, so it is not unitary apart from the trivial one with $t=0$\,. Hence, all partially-massless fields with $t\neq 0$ are non-unitary in AdS space.

Note that the partially-massless field $\cD_{\mathfrak{so}(2,2)}(s-t,s)$
does not have any bulk DoF (as one of two $\mathfrak{so}(2,1)$'s
has a finite-dimensional representation $R_{t/2}$)
but it has $2(t+1)$ boundary DoFs. 
On the other hand,
the gauge mode $\cV_{\mathfrak{so}(2,2)}(s+1,s-t-1)$
has two bulk DoFs as it has infinite-dimensional representation 
for both of $\mathfrak{so}(2,1)$\,.
The maximal-depth case with $t=s-1$ is special here.
Even though the partially-massless field $\cD_{\mathfrak{so}(2,2)}(1,s)$
follows the same pattern as generic $t$\,,
its gauge mode
is given by \emph{two copies} of a parity-invariant scalar mode, 
 $(V_{(s+1)/2}\otimes V_{(s+1)/2})^{\otimes 2}=[\cV_{\mathfrak{so}(2,2)}(s+1,0)]^{\otimes 2}$\,.
This particularity of the maximal-depth
can be understood as well from simple field-theoretical consideratons.

For better understanding, consider the simplest example of spin-one particle. The maximal-depth partially-massless field coincides with the massless field for spin-one particle.
The massive spin-one field is usually described by Proca action,
\be
	S_{\rm\sst Proca}[A]=
	\int d^{3}x\sqrt{g}\left(
	{1 \over 4 e^2} \,F_{\m\n}\,F^{\m\n}+\frac {m^{2}}2\,A_{\m}\,A^{\mu}\right).
	\label{Proca}
\ee
In three dimensions, 
it can also be described as two copies of a \emph{self-dual massive} action 
\cite{Townsend:1983xs},
\be
	S_{\rm\sst SDM}[A^{\pm}]=
	\int d^{3}x\left(
	\frac12\,\epsilon^{\m\n\r}\,A^{\pm}_{\m}\,\partial_{\n} A^{\pm}_{\r}
	\pm\sqrt{g}\,\frac {me}2\,A^{\pm}_{\m}\,A^{\pm\mu}\right),
	\label{SD}
\ee
where $A^{\pm}$ separately describe the $\pm$ helicity modes.
In the massless limit, both actions acquire gauge symmetries
and lose DoFs. 
\begin{itemize}
\item
The Proca action \eqref{Proca}
acquires one gauge symmetry removing 
only one mode $\cV_{\mathfrak{so}(2,2)}(2,0)$ 
from $\cV_{\mathfrak{so}(2,2)}(1,1)$ \eqref{all modes}, 
leaving $\cD_{\mathfrak{so}(2,2)}(1,1)\oplus\cV_{\mathfrak{so}(2,2)}(2,0)$\,.
Hence, together with two boundary modes, it also describes 
a bulk scalar mode.
\item The self-dual action \eqref{SD}
acquires two gauge symmetries:
one for $A^{+}$ and the other for $A^{-}$. Each removes $V_{\frac{s+1}2}\otimes V_{\frac{s+1}2}$, so we end up with two copies of Abelian Chern-Simons, describing $\cD_{\mathfrak{so}(2,2)}(1,1)$\, with only boundary degrees of freedom.

\end{itemize}

In the higher-spin cases,
one can still construct a one-derivative self-dual action
for a massive field that is parity odd. For $t\neq s-1$ partially-massless field,  there is also an equivalent two-derivative parity preserving description. For $t=s-1$ the situation is similar to Maxwell field: the one-derivative description is not equivalent to the two-derivative one.
In the partially-massless limit (including $t=s-1$) of self-dual action, their gauge symmetry
eliminates both $V_{s-\frac t2}\otimes V_{\frac t2+1}$
and $V_{\frac t2+1}\otimes V_{s-\frac t2}$\,,
so completely removes the parity-invariant gauge mode 
$\cV_{\mathfrak{so}(2,2)}(s+1,s-t-1)$
\eqref{gauge mode} \, , 
leaving only the boundary DoF $\cD_{\mathfrak{so}(2,2)}(s+1,s-t-1)$
\eqref{boundary dof}.
On the other hand, beginning with a two-derivative massive action (see e.g. \cite{Zinoviev:2001dt}),
the partially-massless limit attains one \emph{parity-invariant} gauge symmetry.
When the depth is not maximal, $t\neq s-1$\,,
it again removes the parity-invariant combination of gauge modes
$\cV_{\mathfrak{so}(2,2)}(s+1,s-t-1)$\,.
On the contrary, in the maximal-depth case with $t=s-1$\,,
the gauge symmetry removes only one mode
among two   $\cV_{\mathfrak{so}(2,2)}(s+1,0)$'s\,.
Hence, the left-over DoFs
$\cD_{\mathfrak{so}(2,2)}(1,s)\oplus\cV_{\mathfrak{so}(2,2)}(s+1,0)$\, contain a bulk scalar.

The colored higher-spin gravity of $\mathfrak{gl}_M\oplus\mathfrak{gl}_M$ makes use of the self-dual description for (partially-)massless fields.
Hence, the maximal-depth partially-massless field of spin-$M$ 
analyzed in Section \ref{sec: spectrum} does not carry a bulk DoF
but $2\,M$ boundary DoF 
(that is, $M$ left-moving and $M$ right-moving DoF).
This number matches with  
the total boundary DoF of 
massless spin 
$1,2,\ldots,M$.

\subsection*{$dS_3$ case}
In the dS$_3$ space, 
the isometry algebra is given by
$\mathfrak{so}(1,3)\simeq\mathfrak{so}(3)\oplus\overline{\mathfrak{so}(3)}$\,,
and we begin with the representations of 
$\mathfrak{so}(3)$ and $\overline{\mathfrak{so}(3)}$\,.
Differently from the $AdS_3$ case, we do not assume that 
these representations are of lowest-weight type
because in dS we do not have an invariant notion of energy 
 to which we can impose a bound condition.
Still, the representations can be labelled by 
$\mathbb C$ numbers $h$ and $h^{*}$ (for now
$h^{*}$ is different from the complex conjugate $\bar h$)
which parameterize the  Casimir operators of
$\mathfrak{so}(3)$ and $\overline{\mathfrak{so}(3)}$ as
\be
	C=h(h+1)\,,\qquad C^{*}=h^{*}(h^{*}+1)\,.
\ee
From the compactness of $\widetilde{SO(3)}\simeq SU(2)$ in 
$\widetilde{SO(1,3)}\simeq SL(2,\mathbb C)$\,,
we get the quantization condition:
\be
	h-h^{*}=s\in \tfrac12\,\mathbb Z\,,
\ee
which is related to the spin of a particle in dS. 
For convenience, let us define the other combination of $h$ and $h^{*}$ as
\be
	h+h^{*}+1=\mu\,, 
\ee
and $\mu$ is an arbitrary complex number for the moment.
Since $\overline{\mathfrak{so}(3)}$
is the complex conjugate of $\mathfrak{so}(3)$,
for unitarity,
their representations should also be related by 
complex conjugate:
\be
	\bar C=C^{*}\quad
	\Leftrightarrow \quad
	(\bar h-h^{*})(\bar h+h^{*}+1)=0\,. 
\ee
There are two options to satisfy this unitarity condition:
\ba
	 \bar h+h^{*}+1=0\quad &\Rightarrow& \quad 
	 {\rm any}\ s\,,\quad \mu\in i\,\mathbb R\,,
	 \label{generic} \\
	 \bar h-h^{*}=0 \quad &\Rightarrow& \quad 
	s=0\,, \quad \mu\in \mathbb R\,.
	\label{special scalar}
\ea
The first case \eqref{generic} corresponds to the 
usual massive spin-$s$ representation with mass-squared given by 
$\mu^{2}$\,. 
The second case \eqref{special scalar}
corresponds to the special mass region only allowed for the scalar.\footnote{In the complete analysis \cite{HC}, the unitary condition  further restricts the allowed value of $\mu$ to  $|\mu|\le 1$\,.}
Since the representation space 
does not develop any invariant subspace in both of the cases (\ref{generic}, \ref{special scalar}), we do not find any unitary short representation in dS background. In a sense, this is consistent with the fact
that dS space does not have any Lorentzian boundary
where the short representation can live.


\section{Discussions}
\label{sec: discussion}

In this paper, we have analyzed
the theory of colored higher-spin gravity in three dimensions.
We showed that this theory can be viewed as 
a theory of higher-spin gravity and Chern-Simons gauge fields
coupled to matter fields consisting of massless higher spins.
The matter fields introduce multiple saddle point vacua 
with different cosmological constants to the theory, exactly like in the case of 
(A)dS$_{3}$ colored gravity \cite{CS}.
On each of these vacua, the gauge symmetry breaking takes place
which affects the spectrum of the theory.

The mechanism of gauge symmetry breaking and the resulting spectrum 
are interesting. 
First, the Goldstone modes, which are spin-two fields corresponding
to the broken part of the color symmetry,  
are not simply eaten by one of the other fields but 
by "all" other fields. In a sense, it is more correct to describe this as if 
Goldstone modes devours all other spectrum so that they combine altogether to become a single
irreducible Regge trajectory of a maximal-depth partially-massless field.

The nature of partially-massless field is also intriguing.
For the algebra $\mathfrak{gl}_{M}\oplus\mathfrak{gl}_{M}$\,,
it is the spin $M$ maximal-depth partially-massless field, which contains 
all the modes of massless spins from 1 to $M$\,.
In other words, it behaves almost like a massive spin $M$ field
but lacks only one DoF, the scalar mode. 
However, in three dimensions, all (partially) massless fields
with spin greater or equal to one (considering Chern-Simons as spin one)
do not have propagating DoF. They still have boundary modes.
The scalar mode is special as it 
is the only propagating DoF in the three dimensional bulk.
Interestingly, when considering a generic $hs(\l)$ rather than
$\mathfrak{gl}_{M}$, 
we do not have any bound on the highest spin, suggesting that
the maximal-depth partially massless fields, appearing in
the symmetry-broken phase of colored higher-spin gravity, might have
an infinite tower of spin. The entire multiplet is a kind of Regge trajectory, whose slope is set by the nonzero expectation value of the colored spin-2 field and the intercept is set by the depleted spin component state.  
We believe these states are better described when formulated in the Prokushkin-Vasiliev theory
(they correspond to the so-called twisted sector). In the forthcoming papers, we shall study the color decoration of Vasiliev equations in various dimensions as well as several generalizations of the color-decoration mechanism.

\acknowledgments
We thank Marc Henneaux, Jaewon Kim, Jihun Kim, Sasha Polyakov, Mikhail Vasiliev  and, especially, Augusto Sagnotti for many useful discussions. SG and SJR acknowledge the APCTP Focus Program``Liouville, Integrability and Branes (11)" for excellent work environment during this work. EJ thanks the Lebedev Institute for hospitality while this work was being completed. SJR also thanks Lars Brink, the organizer of the  International Workshop on ``Higher Spin Gauge Theories'' at Singapore, for invitation and providing for stimulating environment during this work.  This work was supported in part by the National Research Foundation of Korea through the grant NRF-2014R1A6A3A04056670 (SG, EJ), and the grants 2005-0093843, 2010-220-C00003 and 2012K2A1A9055280 (SG, KM, SJR). The work of EJ is also supported by the Russian Science Foundation grant 14-42-00047 associated with Lebedev Institute. The work of KM was supported by the BK21 Plus Program funded by the Ministry of Education (MOE, Korea) and National Research Foundation of Korea (NRF).

\bibliographystyle{JHEP}

\end{document}